\documentclass[aps,apl,preprint,groupedaddress, showkeys,showpacs]{revtex4-1}
\usepackage[utf8]{inputenc}
\usepackage{amsmath,amssymb}
\usepackage{graphicx,float}
\makeatletter
\usepackage{dsfont}
\usepackage{multirow}
\usepackage{esvect}
\usepackage[acronym,nonumberlist,nopostdot]{glossaries}
\usepackage[margin=1in]{geometry}
\usepackage{standalone}
\usepackage{bbm}
\usepackage{mathbbol}
\title{GGT kinetiks}
\makeatother

\usepackage[bottom]{footmisc}

\newcommand{\iu}{\mathrm{i}\mkern1mu}
\usepackage{color}
\usepackage{xcolor}
\usepackage{cleveref}
\let\oldtop\top
\renewcommand{\top}{\oldtop\!}

\newacronym{PBE}{PBE}{population balance equation}
\newacronym[plural=ODEs, firstplural=ordinary differential equations (ODEs)]{ODE}{ODE}{ordinary differential equation}
\newacronym[plural=GFs, firstplural=generating functions (GFs)]{GF}{GF}{generating function}
\newacronym{HPE}{HPE}{hyperbranched polyester}
\newacronym{NMR}{NMR}{nuclear magnetic resonance}
\newacronym{DMF}{DMF}{dimethylformamide}
\newacronym{FFT}{FFT}{fast Fourier Transform}
\newacronym{MC}{MC}{Monte Carlo}
\newacronym{NWO}{NWO}{Netherlands Organisation for Scientific Research}
\newacronym{MD}{MD}{Molecular Dynamics}

\begin{document}

\title{Dynamic Networks that Drive the Process of Irreversible Step-Growth Polymerization}
\author{Verena Schamboeck} 
\author{Piet D. Iedema} 
\author{Ivan Kryven} 
\affiliation{University of Amsterdam, PO box  94214, 1090 GE, Amsterdam, The Netherlands} 

\begin{abstract}
Many research fields, reaching from social networks and epidemiology to biology and physics, have experienced great advance from recent developments in random graphs and network theory. 
In this paper we propose to view percolation on a directed random graph as a generic model for step-growth polymerisation. 
This polymerisation process is used to manufacture a broad range of polymeric materials, including: polyesters, polyurethanes, polyamides, and many others. 
 We link features of step-growth polymerisation to the properties of the directed configuration model, and in this way, obtain new analytical expressions describing the polymeric microstructure.
Thus, the molecular weight distribution is related to the sizes of connected components, gelation to the emergence of the giant component, and the molecular gyration radii to the Wiener index of these components.
A model on this level of generality is instrumental in accelerating the design of new materials and optimizing their properties.
\end{abstract}


\maketitle

\newpage

\section{Introduction}

Within recent years, network theory became an indispensable tool in a broad range of applied sciences ranging from social psychology and epidemiology to transport engineering, biology and physics\cite{newman2003,strogatz2001}. It allows us to study the spreading of rumours and ideas\cite{moore2000,gomez2013}, but also the spreading of diseases within populations\cite{Manlio2016} and cascades of failures in the electricity grid\cite{callaway2000}. These and alike works established the universal language of network science that is valid across disciplines: it is conceivable that a study on the neural network of the  human brain may teach us to better optimise transportation networks in growing cities\cite{bullmore2009}. 
 The beginnings of network science, by which we refer to a combination of graph and probability theories, are commonly linked  with works of S. Milgram and P. Erd\H{o}s\cite{newman2001b}. 
Nonetheless, not of least importance for the foundation of the field played works of J.P. Flory who proposed to use random, graph-like structures to study hyperbranched and cross-linked polymers\cite{flory1941molecular}.
In fact, currently existing theories of hyperbranched polymers are largely based on the early developments of Flory: the modern viewpoint of network theory is only starting to diffuse back into polymer chemistry where this theory has arguably originated\cite{Papadopoulos2018,kryven2018analytic,orlova2018automated,schamboeck2017acrylate,ye2017,yang2017,kryven2016}.
 
Conventional polymer networks are formed by a process called polymerization, during which small molecules bind together by means of covalent bonds and form large molecular structures. The functionality of these molecules is typically limited by the underlying chemistry and the bonds appear symmetric or asymmetric depending on the nature of functional groups that the reactants of the binding reaction bear\cite{odian2004}.
It is mainly due to the variations of their topologies that polymeric materials feature such a broad range of physical properties. One of the most common polymerisation processes is the step-growth polymerisation of multifunctional monomers. This process leads to hyperbranched polymers of disperse sizes and irregular topology that undergo a phase transition in its connectivity structure during the course of the polymerisation\cite{odian2004}. This  transition is closely related to the percolation on networks\cite{moore2000,callaway2000,gomez2013,Manlio2016}. The phase transition is marked by emergence of the gel, that is the giant molecule that spans the whole volume\cite{newman2003,kryven2016emergence}. 
Flory provided simple analytical expressions for the average molecular size in these systems and was first to explain the onset of gelation, but limited himself to monomers of prescribed type $A_n$ or $A_n+B_2$  \cite{flory1941molecular}. Later, Stockmayer presented a formal expression for the whole distribution of molecular sizes\cite{stockmayer1944theory}, however the practical use of this expression is limited due to combinatorial complexity of computations. Durand and Claude derived a more general analytical expression for averages of the molecular size distribution\cite{durand1979}.
Considerable progress has been made for the case of multifunctional monomers of type $A_n$, which feature symmetric bonds \cite{kryven2018analytic,matsoukas2015statistical,lushnikov2017exactly}, whereas among asymmetric multifunctional monomers only monomers of the type $AB_2$ have a known analytical expression for the molecular size distribution as was demonstrated by Zhou et al. \cite{zhou2006distribution}. For these reasons, the search continued resulting in a wave of fast, approximate methods, as in, for example, works of Kryven et al.\cite{kryven2013novel,kryven2015transition}, Wulkow et. al.\cite{muller1997molecular},  Tobita\cite{tobita2016universality}, Hillegers and Slot\cite{hillegers2017step}. Although these methods are computationally fast, the approximate methods are hard to adapt to new polymerisation schemes, and especially the schemes requiring description with multidimensional distributions.

Yet another approach that has been applied to polymer networks only recently, the molecular dynamics (MD) simulation, is especially attractive as it produces very detailed information on the structure of polymer networks\cite{torres2018,izumi2018molecular}. Molecular dynamics simulations are notorious for being computationally expensive, and therefore, limited to small samples and short time scales. In our previous work\cite{torres2018} we have demonstrated on the case of an acrylate polymer featuring predominantly symmetric covalent bonds, that many of the MD-generated network properties can be also reproduced by the configuration model for undirected random networks\cite{newman2001random,kryven2017general}. 
Furthermore, the recent developments in directed configuration models \cite{kryven2016emergence,kryven2017finite} present an opportunity to develop a generic polymerisation framework that will cover asymmetrical bonds as well. The latter, despite posing a more complex mathematical problem, are also more ubiquitous in polymerisation chemistry, and especially in that of hyperbranched  and super-molecular polymers\cite{metri2018,yan2016,deGreef2008}.

The current paper presents a new look on exactly solvable expressions for hyperbranched polymers by utilising latest developments of random graph models \cite{molloy1995critical,newman2001random,newman2003}, although they might appear still somewhat exotic to the field of polymer chemistry.
Being inspired by the kinetic theory of Krapivsky, Redner, and Ben-Naim \cite{Krapivsky2001,ben2005,ben2011}, we employ a two-stage approach: 
we first devise a kinetic model for the transformations the monomer units undergo in time, 
and then we construct a configuration random graph, which deduces the global properties of the network from the two-variate degree distribution that is obtained on the first stage. 
 Analytical expressions are obtained for various distributional properties of the polymer resulting from step-growth polymerisation of arbitrary combination of arbitrary functional monomers. 
The advantages of the proposed random graph model are grounded in the generic applicability and analytical expressions that are also fast to compute.




%







\section{Results}
\label{sec:results}

\begin{figure}
\includegraphics[width=14cm]{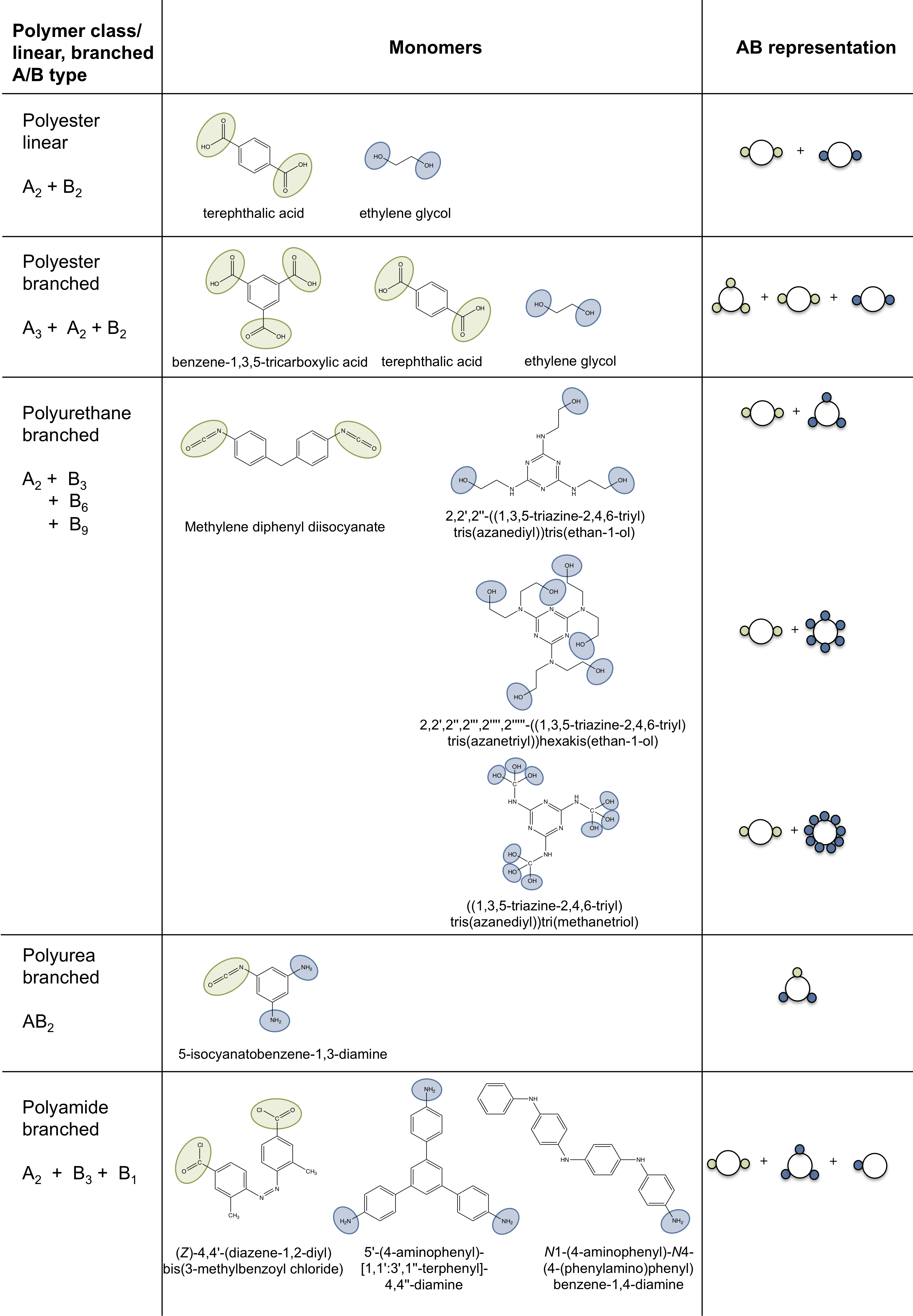}
\caption{Structural formulas for a few examples of linear and branched monomers together with their AB representations that define the underlaying network topology. The list of polymers include: polyester\cite{mckee2005branched}, branched polyurethane\cite{mahapatra2013highly}, polyurea\cite{tuerp2015dendritic} and polyamide\cite{chao2013multifunctional}. }
\label{fig:tab_pol}
\end{figure}

A polymer is a large molecule that consists of many repeat units, the monomers, and is formed as a result of chemical reactions that lead to covalent bonding between the monomers. Step-growth polymerization does not require an initiator and occurs between monomers that carry reactive functional groups. 
Many polymers with real world applications are formed as a result of step-growth polymerisation. Figure~\ref{fig:tab_pol} features a few important examples related to polyesters, polyamides, and polyurethanes. 
The maximum number of chemical bonds that a single monomer bears is limited by the number and type of functional groups that are present in this monomer.
If a system consists of solely two-functional monomers, only linear polymers are  formed in the course of the step-growth polymerisation. 
However, if some (or all) monomers have more than two functional groups, it is possible to form hyperbranched polymers and networks. 

\begin{figure}[t]
\includegraphics[width=7cm]{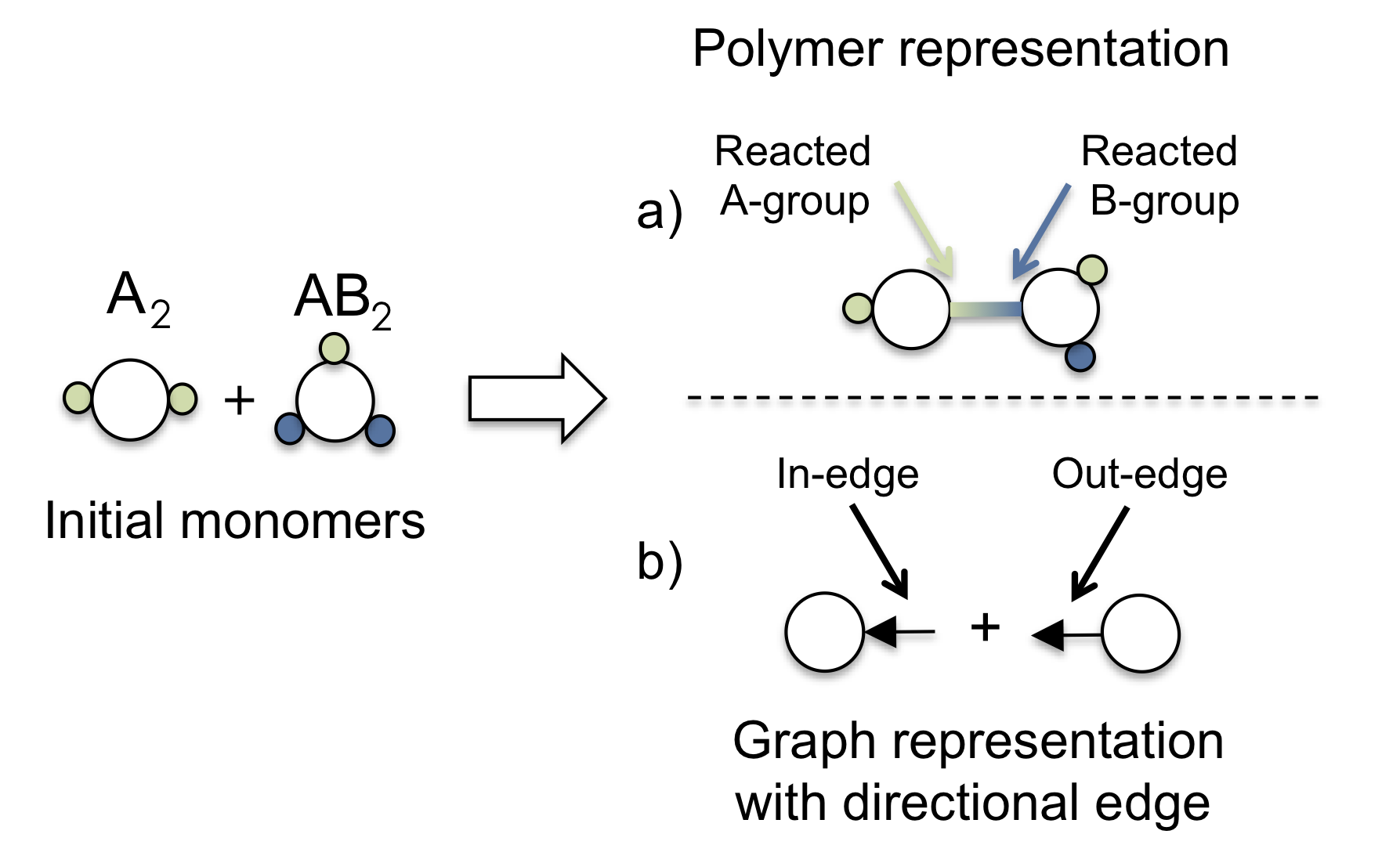}
\caption{
Illustration of an AB-reaction binding one A-group of an A$_2$ and one B-group of an AB$_2$ monomer. In a) the colour of the bond indicates which monomer provided the A- and the B-group. In b), the graph representation, the type of the functional group is stored in the directionality of the edge. An in-edge corresponds to a reacted A-group, an out-edge to a reacted B-group.}
\label{fig:AB_dir}
\end{figure}

In many chemical systems, two monomers bind through an asymmetric reaction that occurs between functional groups of different kinds.
When two functional groups of different kinds are reacting, for example, as in the reaction between an acid and an alcohol leading to an ester, we refer to one group as the A-group, and the other -- as the B-group.
This asymmetric reaction is at the main focus of the current paper.
Symmetric reactions occurring between two groups of the same kind, e.g. two alcohols reacting to form an ether, have been covered elsewhere\cite{kryven2018analytic}.
 Figure~\ref{fig:tab_pol} exemplifies this notation on a few cases of polymers that feature linear and branched topologies.
 For each case, we indicate the structural formulas of the relevant monomers, highlight the functional groups, and give the corresponding AB notation. 
 In Figure~\ref{fig:AB_dir} the asymmetric reaction between A and B groups is illustrated on the example of an $\text{A}_2$ monomer (a monomer with two A-groups) reacting with an $\text{AB}_2$ monomer (a monomer with  one A-group and two B-groups). 

Before introducing the random graph model, we briefly summarise the terminology commonly used in graph theory. A directed graph consists of nodes and directed edges connecting them. 
A subgraph of a graph, in which any two nodes are connected by an undirected path is called a weakly connected component. 
In this work, we drop the prefix weakly, and  refer to this components as  connected components.
 When representing a polymer system as a graph, the monomers are identified with nodes, the chemical bonds with edges and thus a polymer molecule with a whole connected component.
 As the two sides of a chemical bond in the AB-reaction are not identical, we represent this asymmetry with directed edges.
 Without loss of generality, the directionality is defined as pointing from the B-group towards the A-group. The graph representation and the mapping of a reacted A-/B-group to an in-/out-edge is depicted in Figure~\ref{fig:AB_dir}.

\begin{figure}[t]
\includegraphics[width=14cm]{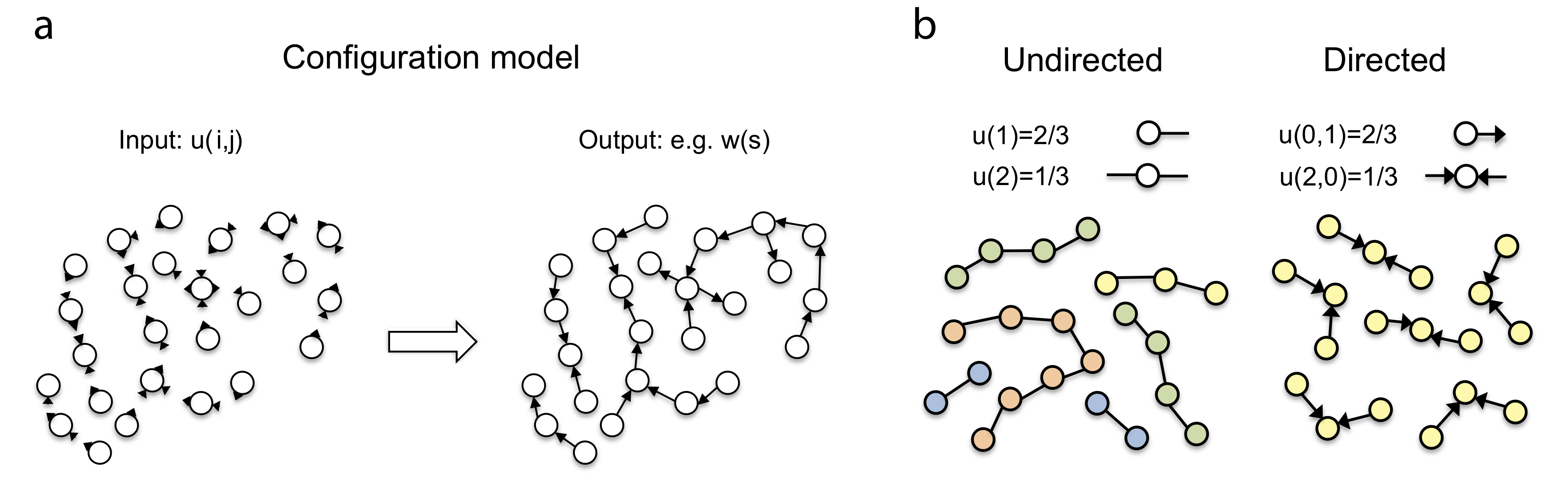}
\caption{
{\bf a,} The concept of the directed configuration model: (\emph{left:}) nodes are signed "half-edges" according to the bivariate degree distribution that provides an input for the model; (\emph{right:}) these nodes are connected randomly so that the degree distribution is strictly satisfied. 
{\bf b,}An example of a trivial degree distribution and the corresponding ensemble of connected components for the case of undirected and directed graphs. }
\label{fig:asymm}
\end{figure}

The number of bonds that a monomer is bearing equals to the degree of the corresponding node. 
Generally speaking, monomers with distinct numbers of bonds may have different concentrations. 
To capture these differences, we refer to the node degree distribution, $u(i,j)$, which defines the probability of a randomly chosen node to have $i$ adjacent in-edges, and $j$ adjacent out-edges, and therefore, $u(i,j)$ is proportional to the concentration of this monomer. 
 Figure~\ref{fig:asymm}b, demonstrates that the directed degree distribution $u(i,j)$ and the projected undirected degree distribution $u(k)=\sum\limits_{i+j=k}u(i,j)$ that ignores the direction of the edges may lead to different sizes for connected components.
In this extreme example, the directed degree distribution only allows connected components of size $s=3$, whereas the  undirected degree distribution does not limit this sizes at all.
It turns out that much of the global information about the polymer system can be deduced from the degree distribution. 

Hyperbranched molecules appear in a broad range of topologies that result from specific chains of reactions, and therefore, may occur with very different frequencies. 
To capture this variability we employ the configuration model that takes a directed degree distribution as an input. In this approach, 
the degree distribution reflects the state of the polymer system as driven by the chemical kinetics, and therefore, this distribution is time-dependant. 
The configuration model maximises entropy of all possible configurations that satisfy a given degree distribution at a time point of interest. This means that the model
is egalitarian with respect to functional groups: every pair of functional groups has equal probability to establish a bond. This principle is illustrated in Figure~\ref{fig:asymm}a.
 The output of the configuration model is then processed to obtain various global properties of the polymer network, as for instance,  the molecular weight distribution (probability that a randomly chosen node belongs to a weakly connected component of size $n$), the gel fraction (probability that a randomly chosen node belongs to the giant component), the mean-square gyration radii (related to the Wiener index of connected components), and the length of the average shortest path. 
The mathematical derivations of these results are presented in the Methods (see Section~\ref{sec:methods}).

\subsection{Reaction kinetics and degree distribution}

The evolution of the degree distribution is governed by the reaction kinetics of the step-growth polymerization. 
This process converts an arbitrary pair of  A- and B-groups into a bond, which is represented as an edge between nodes in the model.
 Since a monomer may carry multiple A- and B-groups, the reaction between a pair of monomers is dependant on the number of unreacted functional groups they carry.
  The formal mechanism for the reaction between two monomers  is given by:
 \begin{equation}\label{eq:process}
(i,j,I,J) + (i',j',I',J') \xrightarrow{k_{AB}(I-i)(J'-j')} (i+1,j,I,J) + (i',j'+1,I',J'),
\end{equation}
where vector $(i,j,I,J)$ denotes the state of a monomer: $I,J\geq0$ are the numbers of respectively A- and  B-groups on this monomer,  and  $i=0,1,\dots,I$, $j=0,1,\dots J$ denote the number of groups of respectively type  A or B that have already been converted into bonds by the reaction. For each monomer, the indices $(I,J)$ are defined a priori, whereas  $(i,j)$ change over time. The reaction rate is given by the product of the rate constant $k_{AB}$, the number of unreacted A-groups on the first monomer $(I-i)$, and the number of unreacted B-groups on the second monomer $(J'-j')$. The following assumptions are made: (1) the reactivity for any pair of A- and B-groups is equal; (2) the reactivity does not change throughout the process. Let $M_{i,j,I,J}(t)$ be the probability that a randomly chosen monomer has configuration $(i,j,I,J)$ at time $t$. This probability is proportional to the concentration of the monomers. 
The time variation $\frac{\partial M_{i,j,I,J}(t)}{\partial t}$ as governed by the process \eqref{eq:process} is described by the corresponding master equation, see the Methods \ref{ssec:master_eq}. In the general case of more than two distinct functional groups (e.g. A-, B-, C-, D-groups) and several reaction mechanisms (e.g. AB-, CD- reaction), the master equation becomes more complex due to a combinatorial number of monomer species of defined type and state. In that case, automated reaction networks can be applied to algorithmically construct the corresponding master equation.\cite{orlova2018automated}

The temporal degree distribution $u(i,j,t)$ is directly deduced from $M_{i,j,I,J}(t)$ by summating over all monomer types $I,J$. As discussed in Section \ref{ssec:master_eq}, the step-growth process Eq.~(\ref{eq:process}) leads to the following degree distribution $u(i,j,t)$:
\begin{equation}
\label{eq:u_p}
u(i,j,t)=\sum_{I,J\geq0} \binom{I}{i}p_\text{A}(t)^i \left(1 - p_\text{A}(t)\right)^{I-i} \binom{J}{j}p_\text{B}(t)^j \left(1 - p_\text{B}(t)\right)^{J-j}P(I,J).
\end{equation}
This expression is given by the product of binomial distributions for the in- and out-edges, with $p_\text{A}(t)$
 and $p_\text{B}(t)$ being the probability that a random A-/B-group is reacted (also referred to as A-/B-conversion), see  Eqs.~(\ref{eq:mu}) and (\ref{eq:prob}).
 The probability distribution $P(I,J)$ defines the initial concentration of monomer types and is referred to as the monomer functionality distribution. Probabilities $P(I,J)$ provide the sole input to the model.
 
\begin{figure}[t]
\includegraphics[width=14cm]{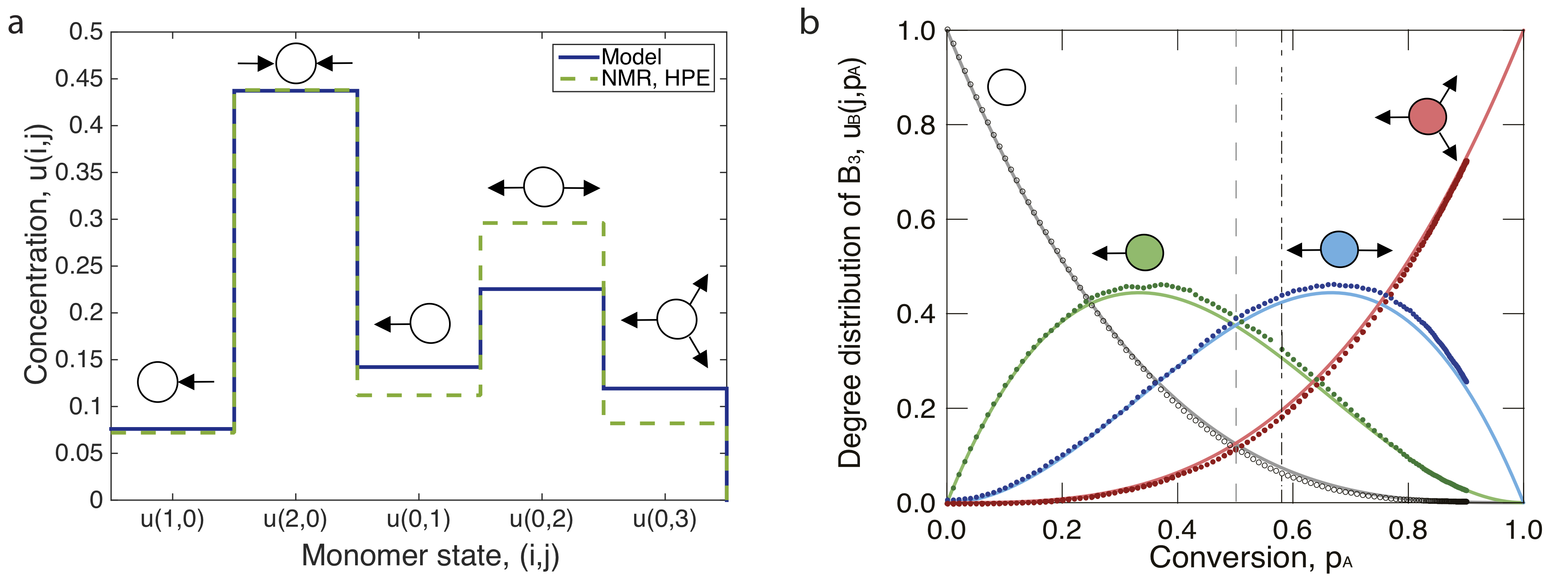}
\caption{
{\bf a,} Comparison of an $^{13}\text{C}$ NMR spectrum (\gls{HPE} in \gls{DMF}\cite{chen2014terminal}) and the predicted degree distribution for the system with $\text{A}_2{:}\text{B}_3=1{:}1$ at $p_A=0.93$.
{\bf b,} Comparison of the phenolic unit degree distribution (only out-edges) in a phenol-methylene system predicted by \gls{MD} (dots) simulation\cite{izumi2018molecular} to the theory (solid lines). The gel point is predicted at $p_A=0.58$ by the \gls{MD} simulation, the theory predicts it at $p_A=0.5$.
 }
\label{fig:NMR}
\end{figure}

In some cases, the degree distribution $u(i,j,t)$ can be measured dirrectly by \gls{NMR}. Due to the chemical shift in the \gls{NMR} spectrum, every monomer state has its own distinct frequency. However, the bigger the monomer is and the more functional groups it has, the harder it is to identify the states. In Figure~\ref{fig:NMR} we compare the degree distributions predicted from the theory with the experimental NMR data from Ref.~\cite{chen2014terminal}, and also, for a different polymerisation system, against \gls{MD} simulation data from Ref.~\cite{izumi2018molecular}.


\subsection{Gelation point and gel fraction}
\label{ssec:gel}
The gelation point marks the transition of the system from a liquid-like to a solid-like state during the polymerization process. 
 After this transition, an increasing positive fraction of monomers becomes a part of single gel molecule. 
 The transition is typically observed by measuring the fraction of the insoluble part of the polymer, or performing rheological experiments.
 From the network theory perspective, this is a well-known phenomenon as the gel transition point corresponds to the emergence of the giant component in the network topology. The size of the giant component is of the order of the whole system size, and  we therefore quantify gel size $g_f$  as the probability that a randomly chosen monomer belongs to the giant component. Section \ref{ssec:gradius} of the Methods explains how $g_f$ can be calculated if the degree distribution is known.
 
The above described process of polymerisation is closely related to percolation on networks. In fact, under certain conditions, the step-growth process is precisely reverse to percolation.
Percolation is typically defined as a process that starts with a full network and removes random edges with given probability.
Let $p$ be the probability that a randomly selected edge is not removed. Under this notation,  percolation can be thought of as a temporal process that starts at $p=0$ and ends at $p=1$. Moreover, this process is known to feature a phase transition at critical probability $p_\text{critical}$, the point at which the giant component appears. When there are equal amounts of A and B functional groups, this process is precisely reverse to the step-growth polymerisation where the edges are being added to a network randomly, and $p_\text{critical}$ coincides with the gel-point conversion of functional groups.
If there are more B-groups than A-groups present in the system initially, the A-groups are limiting the reaction and we conventionally refer to the conversion of A groups as the conversion: $p=p_{A}$. Without loss of generality it may be assumed that the number of A-groups is less or equal to the number of B-groups. 
The moment in time when gelation occurs is completely defined by the proportion of monomers of different functionalities that are present initially in the system.  Generally speaking, the higher the functionality, the earlier gelation occurs in time and/or conversion, and a precise quantitative estimate of the gelation conversion is derived in Section \ref{ssec:gf}.
It turns out that one can connect the critical conversion directly to the functionality distribution $P(I,J)$:
\begin{equation} \label{eq:p_critical}
 	p_{\text{critical}}= \frac{\nu_{01}}{ \nu_{11} +\sqrt{(\nu_{02}-\nu_{01}) (\nu_{20}-\nu_{10})}}.
\end{equation}
where $\nu_{mn}=\sum\limits_{I,J\geq0}I^m J^n P(I,J)$.
Equation \eqref{eq:p_critical} allows one to screen a vast number of systems and determine their gel-point conversions if such occur.
We also can deduce from the theory preset in Methods, Sec. \ref{ssec:gf} that some systems will never form gel. Here again, one can identify whether a system of given monomer functionalities gelates by studying $P(I,J)$. Namely, a monomer system forms gel if the following inequality is satisfied:
$$(\nu_{02} - \nu_{01}) (\nu_{20} - \nu_{10}) - (\nu_{11} - \nu_{01})^2>0.$$

\begin{figure}[t] 
\includegraphics[width=9cm]{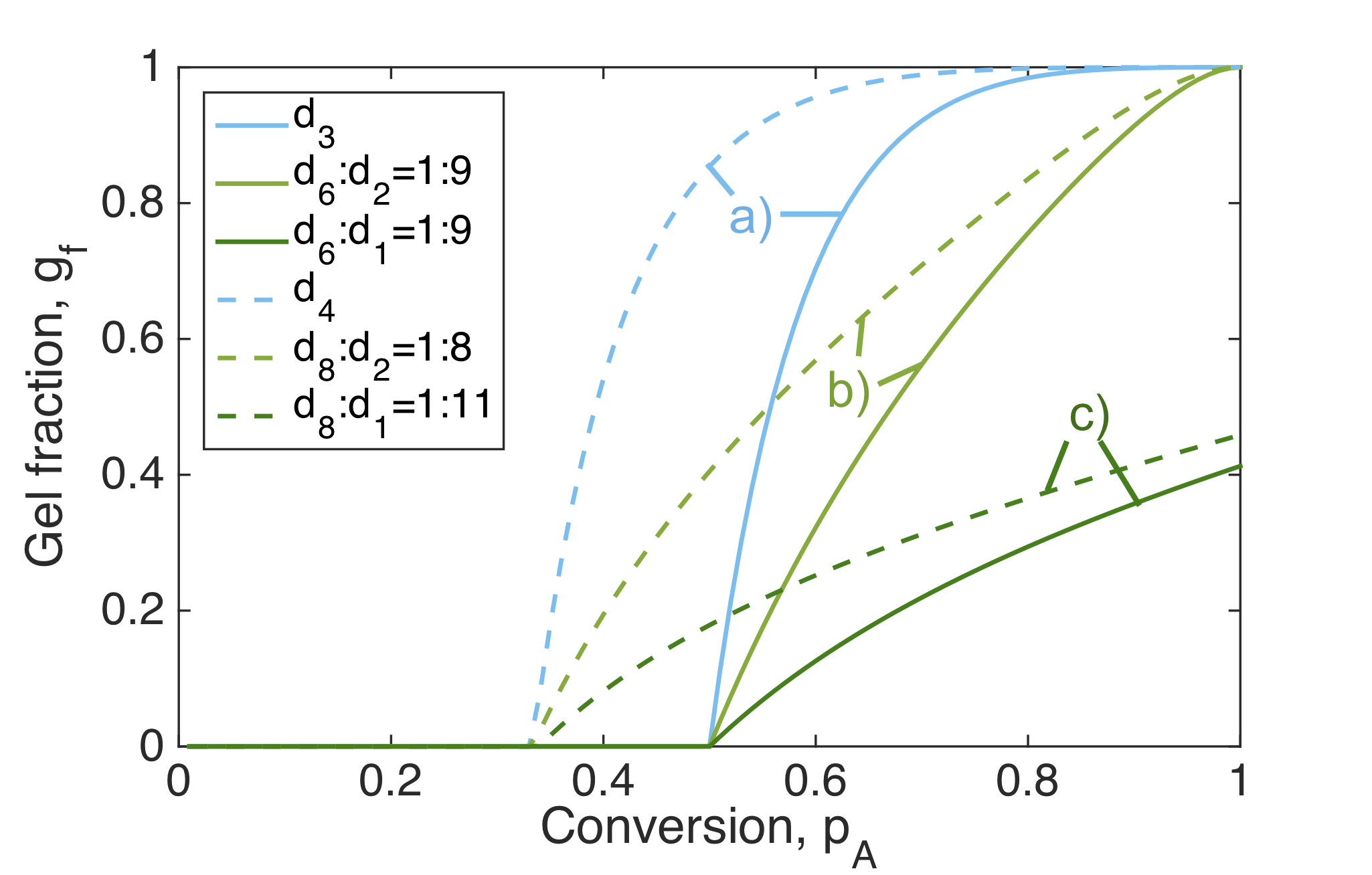}
\caption{ Illustration of 3 types of gel growth behaviour: (a) steep growth, (b) slow growth, (c) gel does not reach full system size. The different types of behaviour are cased by the composition of monomers. Two groups are depicted: (1) solid lines, gel point at $p_{\text{A},\text{critical}}= 0.5$, (a) $A_3{:}B_3=1{{:}}1$, (b) $A_6{:}B_6{:}A_2{:}B_2=1{:}1{:}9{:}9$, (c) $A_6{:}B_6{:}A_1{:}B_1=1{:}1{:}9{:}9$; (2) dashed lines, gel point at $p_{\text{A},\text{critical}}= 0.33$, (a) $A_4{:}B_4=1{:}1$, (b) $A_8{:}B_8{:}A_2{:}B_2=1{:}1{:}8{:}8$, (c) $A_8{:}B_8{:}A_1{:}B_1=1{:}1{:}11{:}11$.}
\label{fig:gf}
\end{figure}

For the physical properties of the final material, both factors  play a definitive role: when does gelation start and how does the gel fraction evolve in the course of the polymerisation? 
 The growth rate of the gel fraction, $g_f$, is determined by the functionalities of the initial monomers and their concentration distribution. In Figure \ref{fig:gf}, a few examples illustrate different types of behaviour of the gel buildup. In these examples, we optimise the initial functionalities and concentrations of monomers to reach two final target properties: (1) a fixed gel point conversion of either $p_{\text{A},\text{critical}}= 0.5$, as depicted by the solid lines, or $p_{\text{A},\text{critical}}= 0.33$, as indicated by the dashed lines;  (2) we distinguish three different types of growth behaviour: (a) a steep growth with most monomers being incorporated into the gel rapidly after gel point, (b) a slow growth with the gel reaching full size only at full conversion, (c) a slow growth with the gel never reaching the system size.
 Behaviour (a) is observed for systems with purely high-functional monomers, (b) for systems with few high-functional monomers and many 2-functional monomers, and (c) for few high-functional monomers and many 1-functional monomers that act as terminal units. The reason for the gel in (c) never reaching the full system size is the formation of small connected components that stop growing because of having all functional groups being capped with one-functional terminal units. For example, when a component is composed of one 6-functional monomer connected to six 1-functional monomers.
 Section \ref{ssec:gf} gives the general equations for the gel-point conversion and gel fractions. 

These results are also interesting when studying polymer ageing and degradation.
Consider a degradation process under which every chemical bond dissociates independently with equal probability. This process is reverse  to the introduced polymerization process, and the gel fraction is a measure of how strongly the system is interconnected.
Clearly,  the systems of type (a) will show a different behaviour during degradation than systems of type (b).
   For (a), the system will stay connected for a long time, but will eventually collapse into many small pieces quite abruptly.
    Type (b) systems will show a more continuous, and therefore more predictable, degradation behaviour, which is also more desirable from applications point of view.

\subsection{Molecular size distribution, averages and asymptotes}
\label{ssec:weight}
From network theory perspective, a separate polymer molecule is a connected component. The sizes of the latter are typically characterised by a size distribution. There are two common ways that such distributions can be defined. The molecular weight distribution $w(s)$ corresponds to the probability that a randomly chosen monomer belongs to a connected component of size $s$, whereas the molecular size distribution $n(s)$ is the probability that a randomly chosen component has size $s$. One can be converted into the other by an appropriate weighting and normalisation: $n(s) = C s^{-1} w(s)$.
In Section~\ref{ssec:ws} we present an exact equation that connects the degree distribution with the molecular weight distribution $w(s)$.
As a general rule, the exact values of $w(n)$ can be computed spending $\mathcal O (n \log n)$ multiplicative operations, and in a special case of only one initial monomer type, the analytic expression for the molecular weight distribution is given in Section~\ref{ssec:1mon}. 

\begin{figure}[t]
\includegraphics[width=\textwidth]{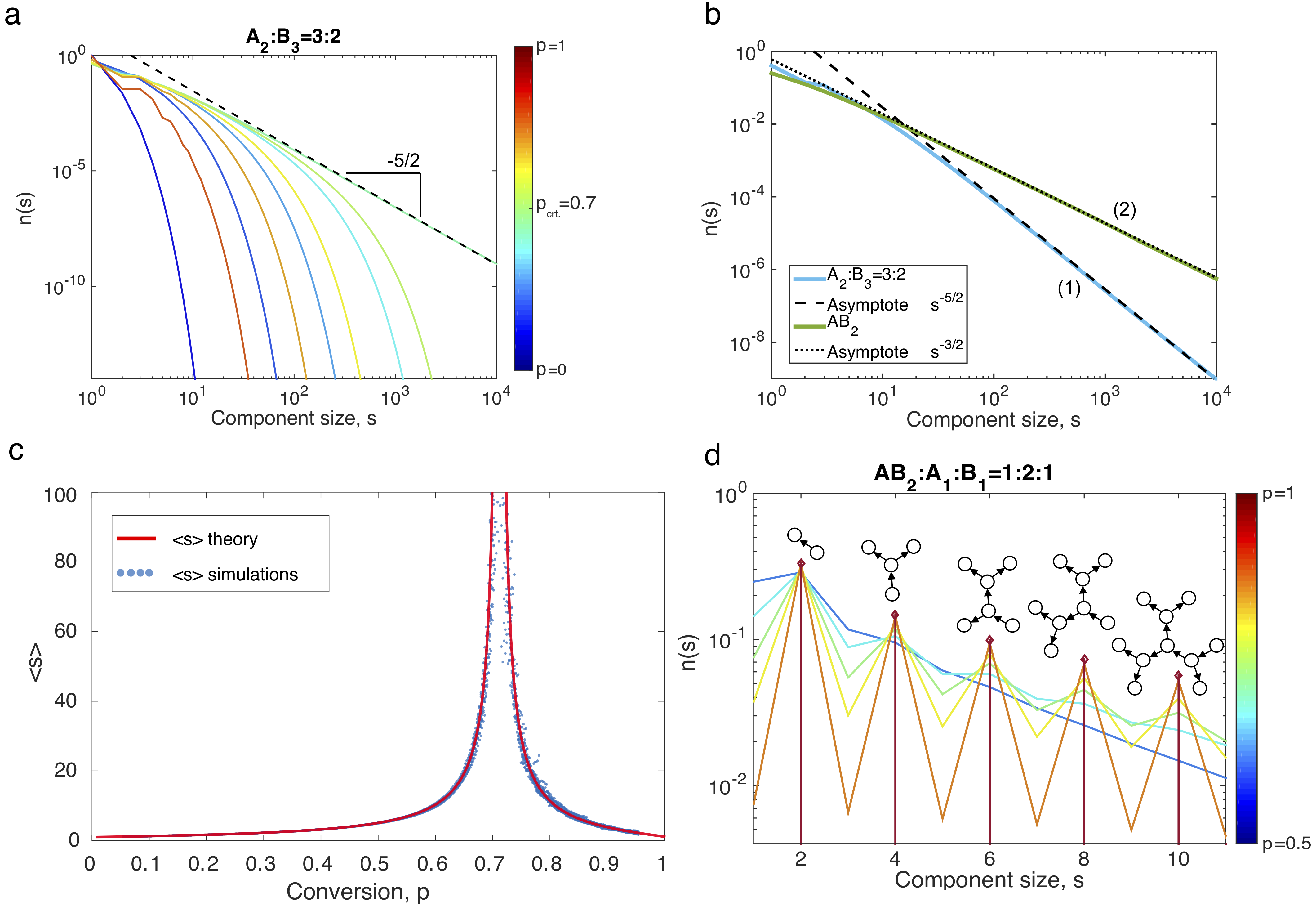}
\caption{
{\bf a,}  Molecular size distributions of the system $\text{A}_2{:}\text{B}_3=3{:}2$ at different conversions as indicated by the colorscale.
 At $p_{\text{A},\text{critical}}=0.707$ the distribution is scale-free, $n(s)\propto s^{-5/2}$.
 {\bf b,} Molecular size distributions $n(s)$  and their critical asymptotes as obtained for two systems: (1) $\text{A}_2{:}\text{B}_3$=3{:}2 and (2) $\text{AB}_2$.
{\bf c,} Weight-average molecular weight for the system $\text{A}_2{:}\text{B}_3=3{:}2$ features singularity at $p_{\text{critical}}=0.7$.
{\bf d,} Oscillating size distributions of the system $\text{AB}_2{:}\text{A}_1{:}\text{B}_1=1{:}2{:}1$ at different conversions as indicated by the colorscale. 
 The examples of dominant polymer topologies are indicated  for reference.}
\label{fig:size_distr}
\end{figure}

The global behaviour that is observed in all polymerising systems can be summarised as follows: Initially, all monomers are unconnected, thus only molecules of size $s=1$ are present.
 With increasing conversion $p_\text{A}$, larger molecules emerge.   The size distribution features the exponential decrease at the tail, and becomes broader with progressing conversion until the gel point $p_{\text{A}\text{,critical}}$ is reached. Only at this single point the size distribution becomes scale-free. Figure~\ref{fig:size_distr}a demonstrates this behaviour on an example.  After the gel transition point, the size distribution describes only the soluble part of the system, so that the size of the gel is given by the gel fraction $g_f=1-\sum\limits_{s=1}^\infty w(s)$. Furthermore, the size distribution returns to its exponential behaviour and becomes narrower with increasing conversion. 

Surprisingly, there are two distinct types of polymerisation systems featuring different types of asymptotic behaviour. Most of polymer systems feature a size distribution with asymptote 
$$n(s)\propto e^{-C_1n}s^{-5/2}.$$
 This, for instance, includes the $\text{A}_2{:}\text{B}_3=3{:}2$ system as illustrated in Figure~\ref{fig:size_distr}a. 
 However, some polymers may also feature a different asymptotic mode, namely 
 $$n(s)\propto  e^{-C_1'n}s^{-3/2}.$$
  This asymptote arises in all systems of type $\text{AB}_n$, for $n>1$.
   In Figure~\ref{fig:size_distr}b, this peculiar case of asymptotic behaviour is  illustrated by comparing two very similar polymer systems $\text{AB}_2$ and  $\text{A}_2+\text{B}_3$ that yet feature different asymptotic modes.


The gel transition is also noticeable in the evolution of weight average molecular weight $M_w$, which features a singularity at the critical point. The evolution of $M_w$ as predicted by the theory is  compared against stochastic simulations in Figure~\ref{fig:size_distr}c. 
The figure shows good agreement between the theory and the numerical simulation except for critical conversion. 
At this point, the stochastic simulations (scatter plot) suffer form the small-system-size effect.
Section~\ref{ssec:av_w} gives analytical equations for the weight average molecular weight in the pre-gel and gel regimes. 


Interestingly, in some cases molecular size distributions feature oscillations. One such example is given in Figure~\ref{fig:size_distr}d, depicting the system $\text{AB}_2{:}\text{A}_1{:}\text{B}_1=1{:}2{:}1$. At full conversion $p_\text{A}=1$ (dark red line) only molecules of specific favoured sizes are present. At lower conversions, also other sizes occur that exhibit strongly reduced probability as compared to the favoured sizes. It turns outs, that the monomers of functionality one play an important role in oscillations, 
as they terminate the growth of polymer molecules and thus fix sizes of these molecules at a constant value. 

\subsection{Gyration radius}
\label{ssec:Rg}
\begin{figure}[t]
\includegraphics[width=0.9\textwidth]{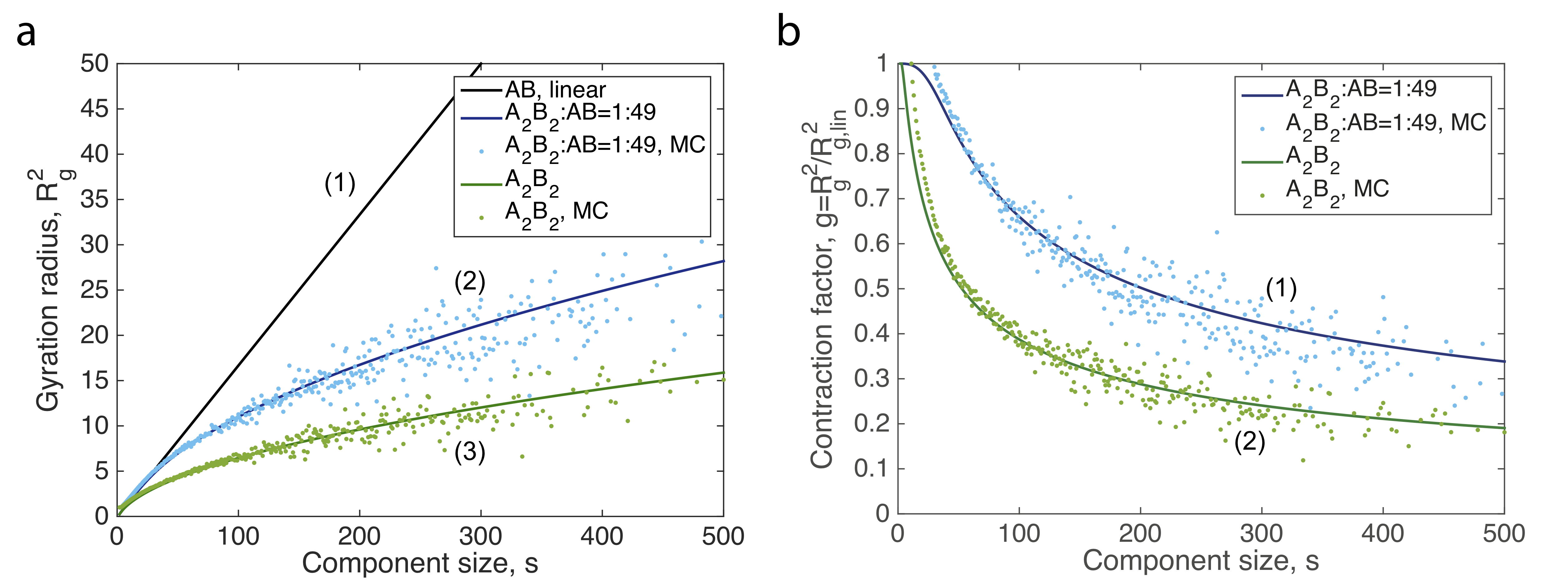}
\caption{ 
{\bf a,}
Theoretical gyration radii (\emph{solid lines}) are compared against simulation data (\emph{scattered data}), which is the average over 100 generated networks that consist of $N=10000$ nodes. 
Three different systems are investigated: (1) linear AB, (2) sparsely branched $\text{A}_2\text{B}_2{:}\text{AB}=1{:}49$, (3) hyperbranched $\text{A}_2\text{B}_2$.
{\bf b,} Theoretical contraction factor of branched components is compared against simulation data.
 Two systems are considered: (1) sparsely branched $\text{A}_2\text{B}_2{:}\text{AB}=1{:}49$, (2) hyperbranched $\text{A}_2\text{B}_2$. }
 \label{fig:gyr_lin}
\end{figure}

Consider a branched polymer molecule that is composed of $s$ monomers. The actual volume this molecule spans is related to how 'branched' this molecule is. 
In systems that contain no gel, or are below the gel transition, it is conventional to characterise this volume by the a quantity called gyration radius $R_\text{g}(s)$, which can also be estimated by light scattering experiments in a polymer solution \cite{burchard1983}.  
 Linear chains feature $R_\text{g,lin}(s)=b\sqrt{\frac{s}{6}}$, where $b$ is the Kuhn length.
 In Section~\ref{ssec:gradius}, we derive the analytical equation that links the degree distribution and the mean square gyration radius for $s\gg1$.
Figure~\ref{fig:gyr_lin}a shows how one can influence $R_\text{g}(s)$ by tuning the set of initial monomers.
 This figure also compares the theoretical gyration radii against gyration radii obtained form stochastically generated networks.
An alternative way of looking at the gyration radius is the contraction factor $g(s)=R_\text{g}^2(s)/R_\text{g,lin}^2(s)$\cite{tacx2017simulating} as displayed in Figure~\ref{fig:gyr_lin}b.
 The contraction factor tells us how much more compact the actual molecule is in comparison to a linear one having the same number of monomers.

Another unexpected result that is revealed by directed random graph theory, is that in step-growth systems all molecules of a fixed size $s$ feature a mean gyration radius that is constant over time.
Since the size distribution does indeed change in time, this time-independency is lost if one calculates an average of this quantity over different molecular sizes.
It is likely that other than step-growth polymerisation processes do not feature time invariant gyration radii.

The gyration radius is also directly proportional of to the Wiener index\cite{dobrynin2001wiener}, which is another topological index to characterise branched molecules. The Wiener index $\mathcal{W}(s)$ is defined as the sum of the lengths of the shortest paths between all pairs of nodes of a graph, the monomer units in the molecule. The relation between the mean square gyration radius and the Wiener index is given by $\mathcal{W}(s)=\frac{R^2_\text{g}(s)}{s^2}$.

\subsection{Scaling of the node neighbourhood size and criticism of well-mixing assumption}
\label{ssec:av_path}

Since the radius of gyration  characterises only finite-sized molecules rather than the gel, which is virtually infinite in size, we are in need of devising an additional measure for the degree of connectedness of the gel. 
With this aim, we investigate the number of nodes at distance $l$ from a randomly selected one, which we shortly refer to as $\mathcal B_l$.
Here, we utilise the topological notion of distance: nodes $i$ and $j$ are at distance $l$ if the shortest undirected path connecting these nodes has length $l$. Since the giant component is infinite in size, the larger the distance $l$ the more nodes are incorporated in the volume of a sphere. In fact, we are mainly interested in the way this quantity asymptotically depends on $l\gg1$. Newman derived an expression for a similar quantity for directed paths\cite{newman2001random}, whereas in the polymer context we are interested in the weak sense of connectivity. This means that we consider the shortest paths that does not respect the directionality of the bonds.

An analytical expression for this behaviour is derived in Section~\ref{ssec:av_path}.
We observe that the average path length in the gel exponentially depends on $l$: 
\begin{equation}
\label{eq:explode}
\mathcal B_l = C e^{ l/l_0},\; l\gg1,
\end{equation}
where $C$ and $l_0$ are constants.
Note that structures that reside in three-dimensional Euclidean space should feature a different scaling law:
$$\mathcal B_l = C N^{d_f}, \; $$
where $1\leq d_f\leq3$ is the cluster growth dimension.
The estimate given in Eq.\eqref{eq:explode} is quite a discouraging result, as a network that features an exponential scaling cannot be physically embedded in the Euclidean space under the condition that the nodes are uniformly distributed with constant density.
This unphysical exponential growth of a node neighbourhood is caused by one of the fundamental assumptions made in the random graph model, and many other popular models that do not explicitly track the spatial configuration of the network, which therefore also sufffer from the same criticism.
In fact, we refer here to one of the most commonly used assumptions of a well-mixed system: any two functional groups react with an equal probability irrespective of their location in the topology. In real systems, however, monomers  interact with the rest of the network that may locally hinder them to react.
That said, it is important to note, that the scaling given by Eq.~\eqref{eq:explode} does not hold before, and precisely at, the gel transition, and the well-mixed system assumption may remain a good approximation in these regimes.

\section{Conclusions}
\label{sec:discussion}
This paper employs recent developments in network science to formalise the step-growth polymerisation process as a problem fully described by network generation. In order to do so, we proposed to view the polymer architecture as a directed network that is described by a dynamic degree distribution.
Although the physical connection between monomers by means of covalent bonding is completely symmetrical, the directionality of the edges keeps record of the asymmetry of the chemical reaction that created the corresponding covalent bond. This approach allows a classification and a general treatment of a vast range of real-world polymerisation problems and can be used for optimisation and design of new materials. As a general rule, the parameters of step-growth polymerising systems comprise a high-dimensional parameter space that dictates the reaction kinetics, network structures, and physical properties of the final material. Therefore, it is important to have a fast  way to map the polymerisation parameters to the final topological properties of the polymer network.

We have matched various idioms present in polymer chemistry to corresponding graph-theoretical analogues and indicated how these can be predicted knowing the input parameters of the system by means of analytical expressions. 
For instance, we gave analytical quantifications of the gelation time,  the topological phase transition and the associated to it molecular weight singularity, the molecular size distribution and its asymptotes, the gyration radii of polymers, and the scaling of the monomer neighbourhood size.
Some of these findings also provide an unexpected qualitative insight on chemistry of polymerisation. For instance, we have revealed the existence of two asymptotical modes that appear in the molecular size distribution, the fact that these size distributions might feature a peculiar oscillating behaviour and that the gyration radii in step-growth polymerised molecules are not dependant on time but only on the sizes of these molecules.

Although they were produced to aid polymer chemistry in the first place, these findings are also relevant to a broader network science community as one can view polymerisation as a process that is reverse to percolation. In this way, the paper amounts to understanding of, so far only poorly studied, percolation on directed networks, which turned out to feature a much richer behaviour then what is observed in undirected networks. 

We finalised the paper with a note of caution that is addressed to the whole modelling community of polymer networks.
The network analysis of the node-neighbourhood scaling in configuration model points out the existence of unphysical features that appear after the gel transition. Importantly, these features are not a complicity of our approach but rather an implication of the commonly trusted assumption of chemical systems being well-mixed. This assumption is standard for many modelling methods that do not explicitly track the coordinates of the chemical species in the three-dimensional space, as for instance is the case for the rate equations, Flory-Stockmayer theory, population balance equations, kinetic Monte Carlo, and other methods. We therefore would like to encourage the search of new network models that bring together mutual interaction of the topology and space.

\section{Methods}
\label{sec:methods}

\subsection{Master equation for the degree distribution\label{ssec:master_eq}} 
In this subsection we briefly summarise the theory from Ref.\cite{kryven2016emergence} that allows us to recover the time evolution of the bivariate degree distribution by constructing an analytically solvable master equation.
We distinguish monomer species by counting the numbers of functional groups of both types $I,J$  and the numbers of in- and out-edges $i,j$.
During the progress of polymerisation the functional groups are converted into chemical bonds between the monomers, and the concentration profiles $M_{i,j,I,J}(t)$ obey the following master equation: 
\begin{equation}
\label{eq:balance}
\begin{aligned}
\frac{\partial}{\partial t} M_{i,j,I,J}(t)  =&
  (I - i + 1) (\nu_{01} - \mu(t)) M_{i-1,j,I,J}(t) \\
& +(J - j + 1) (\nu_{10} - \mu(t)) M_{i,j-1,I,J}(t)\\
& -\Big(\,(I - i) (\nu_{01} - \mu(t)) +(J - j) (\nu_{10} - \mu(t) )\,\Big)M_{i,j,I,J}(t).
\end{aligned}
\end{equation}
Since initially, at $t=0$, there are no bonds, the system is completely described by the distribution of functional groups: $M_{i,j,I,J}(0)=P(I,J),\; i,j=0$ and  $M_{i,j,I,J}=0$ for $i>0$ or $j>0$.
The most important information we like to extract from this master equation is the time dependant degree distribution $u(i,j,t)$. The latter is readily obtained by lumping together all monomer species having the same numbers of in- and out-edges:
\begin{equation}
\label{eq:balance}
u(i,j,t)  = \sum\limits_{I,J\geq0} M_{i,j,I,J}(t).
\end{equation}
The master equation \eqref{eq:balance} is a linear differential-difference equation that can be transformed to an analytically solvable system of partial differential equations by applying $\mathcal Z$-transform.
We thus directly proceed by writing the expression for the degree distibution:
\begin{equation}
\label{eq:u}
u(i,j,t)=\sum_{I,J\geq0} \binom{I}{i}p_\text{A}(t)^i \left(1 - p_\text{A}(t)\right)^{I-i} \binom{J}{j}p_\text{B}(t)^j \left(1 - p_\text{B}(t)\right)^{J-j}P(I,J).
\end{equation}
where
 \begin{equation}
 \label{eq:prob}
\begin{aligned}
 p_\text{A}(t)= \frac{\mu(t)}{\nu_{10}},\; \text{ and }\;  p_\text{B}(t) = \frac{\mu(t)}{\nu_{01}},
 \end{aligned}
 \end{equation}
denote the fractions of converted in- and out-edges, 
and
\begin{equation}\label{eq:mu}
	\mu(t)=\mu_{10}(t)=\mu_{01}(t)=\nu_{01}-\frac{\nu_{01}(\nu_{01}-\nu_{10})}{\nu_{01}-\nu_{10}e^{t(\nu_{10}-\nu_{01})}}
\end{equation}
is the expected in/out degree. Here, $\nu_{mn}$ and $\mu_{mn}(t)$ refer to the mixed moments of respectively $P(I,J)$ and $u(i,j,t)$:
\begin{equation}\label{eq:mu_nu}
	\begin{split}
	&\nu_{mn}=\sum\limits_{I,J\geq0}I^m J^n P(I,J),\\
	&  \mu_{mn}(t)=\sum\limits_{i,j\geq0}i^m j^n u(i,j,t).
	\end{split}
\end{equation}
Note, that since the expected in- and out-degrees coincide, we have $\mu(t)=\mu_{10}(t)=\mu_{01}(t)$.
Mixed moments of the degree distribution $\mu_{mn}(t)$ can be obtained in the form of analytical expressions by performing the appropriate summations of Eq.~\eqref{eq:u}. For instance, the list of mixed moments up to order $n+m=2$ is as follows:
\begin{equation}\label{eq:moments}
	\begin{split}
		&\mu_{00}(t)=1,\\
		&\mu_{10}(t)=\mu_{01}(t)=\mu(t)=p_\text{A}(t)\nu_{10}=p_\text{B}(t)\nu_{01},\\
		&\mu_{20}(t)=p_\text{A}(t)(p_\text{A}(t)\nu_{20}-p_\text{A}(t)\nu_{10}+\nu_{10}),\\
		&\mu_{02}(t)=p_\text{B}(t)(p_\text{B}(t)\nu_{02}-p_\text{B}(t)\nu_{01}+\nu_{01}),\\
		&\mu_{11}(t)=p_\text{A}(t)p_\text{B}(t)\nu_{11},\\
		&\mu_{22}(t)=p_\text{A}(t) p_\text{B}(t)\Big(p_\text{A}(t) p_\text{B}(t)\big(\nu_{22}-\nu_{21}-\nu_{12}+\nu_{11}\big)+p_\text{A}(t)\big(\nu_{21}-\nu_{11}\big)+p_\text{B}(t)\big(\nu_{12}-\nu_{11}\big)+\nu_{11}\Big).
	\end{split}
\end{equation}
We make use of the expressions for $\mu_{i,j}(t)$ in Section \ref{ssec:gf} to determine the gelation conversion, whereas $u(i,j,t)$  is linked to the distribution of molecular weights in Section~\ref{ssec:ws}, to average molecular weight in Section~\ref{ssec:av_w}, and to the typical shortest path length in Section~\ref{ssec:av_path}.

\subsection{Generating functions of the degree and excess degree distributions}
\label{ssec:gen_func}
The utility of the \glspl{GF} for studying random networks was popularised by Newman and his coauthors. See, for example, Ref. \cite{newman2010}. In this section, we omit the time dependance, where it leads to no confusion, and discuss this theory in the context of directed networks defined by the bivariate degree distribution \eqref{eq:u}. 
The \gls{GF} of a bivariate distribution is formally given  by

\begin{equation}\label{eq:gf_u}
	U(x,y)=\sum_{i,j\geq0}u(i,j)x^i y^j,
\end{equation}
with $|x|,\,|y|\leq 1$, $x,y\in \mathbb{C}$ and $U(x,y)|_{x,y=1}=1$. 
The excess distributions $u_\text{in}(i,j)$ and $u_\text{out}(i,j)$, are defined as the degree distributions of nodes that are reached by randomly choosing an in- or out-edge:
\begin{equation}\label{eq:uall}
\begin{split}
	&u_\text{in}(i,j)=\frac{1}{\mu}(i+1)u(i+1,j),\\
	&u_\text{out}(i,j)=\frac{1}{\mu}(j+1)u(i,j+1).
\end{split}
\end{equation}
The \glspl{GF} of the latter distributions are given by:
\begin{equation}\label{eq:Uall}
\begin{split}
	&U_\text{in}(x,y)=\frac{1}{\mu}\frac{\partial U(x,y)}{\partial x},\\
	&U_\text{out}(x,y)=\frac{1}{\mu}\frac{\partial U(x,y)}{\partial y},
\end{split}
\end{equation}
that satisfy $U_\text{in}(x,y)|_{x,y=1}=1$ and $U_\text{out}(x,y)|_{x,y=1}=1$.

Now plugging the degree distribution \eqref{eq:u} into Eqs.~\eqref{eq:u},\eqref{eq:gf_u}, and \eqref{eq:Uall} gives:
\begin{equation}\label{eq:U_expl}
	\begin{split}
		&U(x,y)=\sum\limits_{I,J\geq0} P(I,J)((x-1)p_\text{A}+1)^{I} ((y-1)p_\text{B}+1)^{J},\\
		&U_\text{in}(x,y)=\frac{1}{\mu}\sum\limits_{I,J\geq0} P(I,J) I p_\text{A}((x-1)p_\text{A}+1)^{I-1}((y-1)p_\text{B}+1)^{J},\\
		&U_\text{out}(x,y)=\frac{1}{\mu}\sum\limits_{I,J\geq0} P(I,J) J p_\text{B} ((x-1)p_\text{A}+1)^{I} ((y-1)p_\text{B}+1)^{J-1}.
	\end{split}
\end{equation}
Having these expressions in hand, the moments of the degree distribution $u(i,j,t)$, defined by Eq.~(\ref{eq:mu_nu}), can be directly linked to the partial derivatives of the \glspl{GF} by
\begin{equation}\label{eq:U_mul}
	\begin{split}
		\mu_{mn}=\left[ \left(x\frac{\partial}{\partial x}\right)^m \left(y\frac{\partial}{\partial y}\right)^n U(x,y) \right]\bigg|_{x,y=1}.
	\end{split}
\end{equation}

These relations are used in Sections \ref{ssec:ws}-\ref{ssec:av_path} to derive analytical expressions for various global features of the polymer network.

\subsection{Molecular weight distribution}
\label{ssec:ws}
From chemical point of view any cluster of monomers that are connected together by means of covalent bonds is considered to be a molecule.
In our directed network a molecule is therefore represented by a connected component, whereas the molecular weight is simply the size of this component. The distribution of molecular weights is a popular descriptor of polymer materials.
In fact, there are two ways to define such distribution: $w(s)$ the probability that a randomly chosen \emph{monomer} belongs to a component of size $s$ is called the molecular weight distribution. 
Alternatively, by applying weight $\frac{1}{s}$ we obtain the molecular size distribution, $$n(s)=\frac{C}{s}w(s),$$ that is the probability that a randomly chosen \emph{molecule} has size $s$. In the latter equation $C$ provides the appropriate normalisation of probability.

Here we link the molecular weight distribution $w(s)$ to the size distribution of connected components in the directed configuration model as derived in Ref.\cite{kryven2017finite}, and briefly discuss the insights this interpretation brings to understanding the step-growth polymerisation polymerisation process.
 The first values of $w(s)$ can be found by following simple considerations. For instance $w(1)$ is the probability to choose an isolated node with no neighbours, and therefore: $$w(1)=u(0,0).$$ Furthermore, $w(2)$ is the probability that a randomly chosen node has one edge and its only neighbour has no edges except the one that connects it with the first node: 
$$w(2)=\frac{2}{\mu}u(1,0) u(0,1).$$ 
Continuing this list would lead to a combinatorial explosion of possibilities. A much faster alternative is to employ the \glspl{GF}.
The \gls{GF} for $w(s)$ is formally defined as: 
\begin{equation}\label{eq:gf_w}
	W(x)=\sum_{s}w(s)x^s,\; |x|\leq 1,\; x\in \mathbb{C},
\end{equation}
 and is obtained from the following system of functional equations:
\begin{equation}\label{eq:Wa}
	\begin{split}
	&W(x) = x U\Big[ W_{\text{out}}(x), W_{\text{in}}(x) \Big],\\
	&W_{\text{in}}(x)  = x U_{\text{in}} \Big[ W_{\text{out}}(x), W_{\text{in}}(x) \Big],\\
	&W_{\text{out}}(x) = x U_{\text{out}}\Big[ W_{\text{out}}(x), W_{\text{in}}(x) \Big],
	\end{split}
\end{equation}
where the \glspl{GF} $U(x,y)$, $U_\text{in}(x,y)$ and $U_\text{out}(x,y)$ are defined by Eqs.~(\ref{eq:gf_u}) and (\ref{eq:Uall}). The functions $W_\text{in}(x)=\sum_{s>0}w_\text{in}(s)x^s$ and $W_\text{out}(x)=\sum_{s>0}w_\text{out}(s)x^s$ denote the \glspl{GF} of the excess component size distributions $w_\text{in}(s)$ and $w_\text{out}(s)$.
 
The formal solution to \eqref{eq:Wa} is given by the following relation \cite{kryven2017finite}: 
\begin{equation}\label{eq:w2d}
	w(s) =
	\begin{cases}
		\sum\limits_{m=0}^{s-1} a(m,s-m-1), & s>1;\\
 		u(0,0), & s=1,
	\end{cases}
\end{equation}
where
\begin{equation}\label{eq:ajk}
	a(m,n)= u(i,j)* u_{\text{out}}(i,j)^{*m-1}* u_{\text{in}}(i,j)^{* n-1 } * d(i,j)\Big|_{\substack{i=m\\j=n}},\;m,n\geq0,
\end{equation}
and
\begin{equation}\label{eq:Djk}
	d(i,j)= [ u_{\text{out}}(i,j)- i u_{\text{out}}(i,j)] *  [ u_{\text{in}}(i,j) - j u_{\text{in}}(i,j) ] - j u_{\text{out}}(i,j) * i u_{\text{in}}(i,j).  
\end{equation}
Here, $f(i,j)^{*n}$ denotes the convolution power $f(i,j)^{*n}=f(i,j)^{*n-1} * f(i,j),\; f(i,j)^{*0}:=1$, and the bivariate convolution is defined as 
\begin{equation}\label{eq:convnd}
	f(i,j)*g(i,j)= \sum\limits_{i_1 +i_2=i, j_1+j_2=j}f(i_1,j_1)g(i_2,j_2).
\end{equation}
In practice, numerical values of the convolution can be conveniently obtained by Fast Fourier Transform (FFT).
The asymptotical analysis of $w(s)$ yields two distinct asymptotes:
\begin{equation}\label{eq:asymptote}
	w_\infty(s) \propto  s^{-\frac{3}{2} }e^{ -C_1 s  },\; \text{if } \nu_{20}>0\; \text{and}\; \nu_{02}>0
\end{equation}
and
\begin{equation}\label{eq:degenerate_eq}
	w_{\infty}(s) \propto  s^{-\frac{1}{2}}  e^{ -C_1' s },\; \text{if } \nu_{20}=0\; \text{or}\; \nu_{02}=0
\end{equation}
The exact expression for the coefficients $C_1$ and $C_1'$ are given in  Ref.~\cite{kryven2017finite} and $\nu_{nm}$ are as defined in Eq.~\eqref{eq:mu_nu}.

\subsection{Systems with a single monomer type}
\label{ssec:1mon}
Although the numerical values of Eq.~\eqref{eq:w2d} are accessible in the cost of $\mathcal O(s^2\log s)$ operations, explicit analytical relations can be obtained in  some special cases. Consider the case when there is only one monomer species bearing $I$ groups of type A and $J$ groups of type B, that is the $\text{A}_I\text{B}_J$ monomer. 
We will now derive an explicit analytical expression for $w(s)$.
Note that in this case, the distribution of initial functionalities is trivial, $P(I,J) = 1$, and therefore
the expression for the degree distribution given in Eq.~(\ref{eq:u}) simplifies to a bivariate binomial distribution:
\begin{equation}
\label{eq:u_1}
	\begin{aligned}
		&u(i,j,t)=\binom{I}{i}p_\text{A}(t)^i \left(1 - p_\text{A}(t)\right)^{I-i} \binom{J}{j}p_\text{B}(t)^j \left(1 - p_\text{B}(t)\right)^{J-j},
	\end{aligned}
\end{equation}
where
\begin{equation}\label{eq:cj}
	p_\text{A}(t)=J \frac{\text{e}^{t (I-J)}-1}{I \text{e}^{t (I-J)}-J}\; \text{and}\;p_\text{B}(t)=  I  \frac{ \text{e}^{t (I-J)}-1}{I \text{e}^{t (I-J)}-J}.
\end{equation}
The first values of $w(s)$ are readily obtained by writing:
$$
\begin{aligned}
&w(1)=u(0,0)=( 1 - p_A )^I  ( 1 - p_B )^J,\\
&w(2) = \frac{2}{\mu} u(1,0)u(0,1) = 2  J p_B   ( 1 - p_A )^{ 2  I - 1 }  ( 1 - p_B )^{ 2 J - 1 }=2  I p_A   ( 1 - p_A )^{ 2  I - 1 }  ( 1 - p_B )^{ 2 J - 1 }.
\end{aligned}
$$
Since $u(i,j)$ has a binomial form, one can analytically solve the convolution powers appearing in Eq.~(\ref{eq:ajk}) to obtain:
\begin{multline}
	w(s)=p_\text{B}^{s-2} (1-p_\text{A})^{(I-1) s+1} (1-p_\text{B})^{(J-1) s+1} \times\\
	\left( p_\text{B} \binom{J s}{s-1} \, A  -p_\text{B} J \binom{J s-1}{s-2} B-p_\text{A} (I-1) \binom{J s-2}{s-2}  C \right),
\end{multline}
where factors $A,B,$ and $C$ are defined via the Hypergeometric function:
\begin{equation*}
	\begin{aligned}
		A=&_2\text{F}_1\!\left(1-s,\,(I-1) s+1;\,\,\,-J s;  \,\, \frac{p_\text{A}}{p_\text{B}}\right),\\
		B=&_2\text{F}_1\!\left(2-s,\,(I-1) s+1;\,\,\,1-J s; \,\, \frac{p_\text{A}}{p_\text{B}}\right),\\
		C=&_2\text{F}_1\!\left(2-s,\,(I-1) s+2;\,\,\,2-J s; \,\, \frac{p_\text{A}}{p_\text{B}}\right).
	\end{aligned}
\end{equation*}

\subsection{Weight-average molecular weight}
\label{ssec:av_w}
The weight-average molecular weight is a widely-used quantity in polymer chemistry.  It is defined by the follwing ratio: 
\begin{equation}\label{eq:av_n1a}
\langle s \rangle:=\frac{\sum_{s>0}sw(s)}{\sum_{s>0}w(s)}=\frac{W'(1)}{W(1)}.
\end{equation}
Note that before the gel point, $w(s),w_\text{in}(s)$ and $w_\text{out}(s)$ are appropriately normalised, $$\sum_{s>0}w(s)=W(1)=W_\text{in}(1)=W_\text{out}(1)=1.$$
 By plugging  definition \eqref{eq:av_n1a} into Eq.~\eqref{eq:Wa} we obtain:
\begin{equation}\label{eq:av_n1}
	\langle s \rangle_{t<t_\text{gel}}=\frac{W'(1)}{W(1)}=W'(1)=\frac{\mu^2(-2\mu_{11}+\mu_{20}+\mu_{02})}{\mu_{11}^2-2\mu\mu_{11}-\mu_{02}\mu_{20}+\mu(\mu_{20}+\mu_{02})}+1.
\end{equation}
After gel transition, the latter expression becomes more complex and reads:
\begin{equation}\label{eq:av_n2}
\begin{aligned}
&	\langle s \rangle_{t>t_\text{gel}}=\frac{W'(1)}{W(1)}=\\
&\frac{\mu^2}{U(r_\text{out},r_\text{in})}  \cdot \frac{ 2 r_\text{in} r_\text{out} (\mu- U_{11}(r_\text{out},r_\text{in}))+r_\text{in}^2U_{02}(r_\text{out},r_\text{in}) +r_\text{out}^2U_{20}(r_\text{out},r_\text{in})}{\mu^2-2\mu  U_{11}(r_\text{out},r_\text{in})-U_{02}(r_\text{out},r_\text{in})U_{20}(r_\text{out},r_\text{in})+U_{11}^2(r_\text{out},r_\text{in}) }+1,
	\end{aligned}
\end{equation}
where 
$
	U_{lm}(x,y)=\bigg(\frac{\partial}{\partial x}\bigg)^l \bigg(\frac{\partial}{\partial y}\bigg)^m U(x,y)
$
 denote the partial derivatives,
and $(r_\text{in},r_\text{out})$ is the solution of the following system of equations:
\begin{equation}\label{eq:r_in_out}
	\begin{split}
	&r_\text{in}=U_\text{in}(r_\text{out},r_\text{in}),\\
	&r_\text{out}=U_\text{out}(r_\text{out},r_\text{in}).
	\end{split}
\end{equation}



\subsection{Phase transition and gel fraction\label{ssec:gf}}
 During the evolution of the network the functional groups are converted into edges and at some critical point of time the system accumulates so many edges that it percolates. This critical moment can be identified by a few alternative methods.
For instance, one may study the asymptotical behaviour of the size distribution of connected components as given by \eqref{eq:asymptote}. This asymptote becomes scale-free at the critical point. The other alternative is to directly detect the percolation phase transition by looking at the degree distribution itself. In this case, the changes that occur in the degree distribution at the critical point are more subtle, yet they can be detected by a specially designed criticality criterion. This criterion was given by Molloy and Reed for undirected networks \cite{molloy1995critical}, and was later generalised to the case of directed networks in Ref. \cite{kryven2016emergence}.  In this section we briefly discuss the implications of the latter theory on our dynamic polymer network.

If the only available information about a system is its degree distribution, we can detect whether the system is in the gel regime by the following criterion:
\begin{equation} \label{eq:gel1}
 	\mu_{11}^2-2\mu\mu_{11}-\mu_{02}\mu_{20}+\mu(\mu_{20}+\mu_{02})\le 0.
\end{equation}

The conversion of A-groups at the critical point is given by: 
\begin{equation} \label{eq:gel}
 	p_{\text{A}, \text{critical}}= \frac{\nu_{01}}{ \nu_{11} +\sqrt{(\nu_{02}-\nu_{01}) (\nu_{20}-\nu_{10})}}.
\end{equation}
Thus, if $p_\text{A}(t)>p_{\text{A},\text{critical}}$ the system contains gel.
Some system, however, never produce gel. 
This happens because the initial configuration of the system does not have a sufficient amount of high functional monomers, or too many monomers of functionality one that terminate the growth of the network. 
In either case this statement can be quantified by looking at the moments of the functionality distribution $P(I,J)$:
the phase transition occurs in finite time if at least one of the following conditions is true:
\begin{equation} \label{eq:criterion2}
	\begin{aligned}
		(\nu_{02} - \nu_{01}) (\nu_{20} - \nu_{10}) - (\nu_{11}& - \nu_{01})^2>0, \text{ and }   \nu_{01} \geq \nu_{10},\\
		\text{or     \hspace{4.8cm}} &\\
		(\nu_{02} - \nu_{01}) (\nu_{20} - \nu_{10}) - (\nu_{11} &- \nu_{10})^2>0,  \text{ and }   \nu_{01} \leq  \nu_{10}.
	\end{aligned}
 \end{equation}

The gel fraction is defined as the probability that a randomly selected node belongs to the gel molecule.
The \gls{GF} of the component size distribution $W(x)$ only describes the components of finite size, and the gel fraction is found as the mass deficit that departs from zero at the phase transition. The amount of this 'lost' mass, that is the probability that a randomly chosen monomer belongs to the gel,  is given by
\begin{equation}\label{eq:g_f}
	g_f=1-r,
\end{equation}
where $r=W(1)$. This means that in order to recover $g_f$, one needs to solve the equation for $W(x)$ only at a single point $x=1$.
By substituting  $x=1$ into \eqref{eq:Wa} one obtains:
\begin{equation}\label{eq:Win_Wout}
	\begin{split}
		&r=U(r_\text{out},r_\text{in}),\\
		&r_\text{in}=U_\text{in}(r_\text{out},r_\text{in}),\\
		&r_\text{out}=U_\text{out}(r_\text{out},r_\text{in}),
	\end{split}
\end{equation}
where $U(x,y)$, $U_\text{in}(x,y)$ and $U_\text{out}(x,y)$ are given by Eq.~(\ref{eq:U_expl}). 


\subsection{Gyration radius\label{ssec:gradius}}
In polymer physics, the radius of gyration is used to describe the dimensions of a branched polymer and can be experimentally observed by light scattering experiments.
Consider a branched polymer  with $s$ monomers having coordinates $r_i\in\mathbb R^3, $ $i=1,\dots,s.$
The radius of gyration $R(s)^2_\text{g}$ of this topology is conventionally defined as:
\begin{equation}\label{eq:R}
 R_\text{g}^2(s)= \frac{1}{s^2}\sum^{s}_{k=1}\sum^s_{l=k}(\vec{r}_k-\vec{r}_l)^2.
\end{equation}
This quantity can be estimated using the Kramer's theorem\cite{rubinstein2003polymer} that states:
\begin{equation}\label{eq:R2_Kram}
	\frac{R_\text{g}^2(s)}{b^2}=\frac{1}{s^2}\sum_{j=1}^{s-1} s_L(j)s_R(j),
\end{equation}
where $b$ is the Kuhn's length, which is related to the size of a monomer unit. This sum runs over all possible cuts of the branched structure into two fragments: the left fragment of size $s_L$ and the right fragment of size $s_R$. There are $s-1$ of such partitions.  In a statistical ensemble, the size distributions of  $s_L$ and $s_R$ are given by respectively $w_\text{in}(s)$ and $ w_\text{out}(s)$, which are defined by their generating functions in Eq.~\eqref{eq:Wa}.
Hillegers \& Sloot~\cite{hillegers2017step} formulated the ensemble average for  Eq.~(\ref{eq:R2_Kram}) with respect to $w_\text{in}(s)$ and $ w_\text{out}(s)$:
\begin{equation}\label{eq:R2}
\begin{split}
	\frac{\langle R_\text{g}^2(s)\rangle}{b^2}&=\frac{1}{s^2}\frac{\sum_{s_A+s_B=s} s_A w_\text{in}(s_A) s_B w_\text{out}(s_B)}{\frac{1}{s-1} \sum_{s_A+s_B=s} w_\text{in}(s_A) w_\text{out}(s_B)},
\end{split}
\end{equation}
which we further process it using a discrete Fourier transform $\mathcal{F}^{-1}G(k)|_s$:
\begin{equation}\label{eq:gyration}
\begin{aligned}
	\frac{\langle R_\text{g}^2(s)\rangle}{b^2}=\frac{s-1}{s^2}\frac{  \Big(sw_\text{in}(s)\Big)*\Big(sw_\text{out}(s)\Big) }{ w_\text{in}(s) * w_\text{out}(s)}=
	\frac{s-1}{s^2}    \frac{ \mathcal{F}^{-1} \Big(   W_\text{in}'(x_k) W_\text{out}'(x_k)\Big) \Big|_{s-2} }  {   \mathcal{F}^{-1} \Big( W_\text{in}(x_k)W_\text{out}(x_k)\Big) \Big|_{s} },	
	\end{aligned}
\end{equation}
with $x_k = e^{ -2 \pi \iu \frac{k}{ N+1}}$, $k=0,\dots,N$.
Therefore, if $W_\text{in}(x)$ and $W_\text{out}(x)$ are already available from the computation of the component size distribution $w(s)$, the ensemble-average radius of gyration is obtained by applying the FFT algorithm the cost of $\mathcal O(s \log s)$ multiplicative operations. 
For small  $s$, it is possible to compute the convolution directly, whereas  Eq. \eqref{eq:gyration} is the most advantageous for $s\gg1$, where the direct evaluation of the convolution becomes unfeasible.

Another related quantity to the gyration radius is the contraction factor $g(s)$, which is given by the ratio between the mean square gyration radius for a branched polymer $R_\text{g}^2(s)$ and the square gyration radius of a reference linear polymer with the same length $R_\text{g}^2(s)|_\text{linear}$:
\begin{equation}\label{eq:shrink}
	g(s)=\frac{R_\text{g}^2(s)}{R_\text{g}^2(s) |_\text{linear}},
\end{equation}
where $R_\text{g}^2(s) |_\text{linear}$ is typically estimated from the Gaussian coil model:
\begin{equation}\label{eq:R2_lin}
	\frac{R_\text{g}^2(s)}{b^2}\bigg|_\text{linear}=\frac{s^2-1}{6s}.
\end{equation}

\subsection{Scaling of the node neighbourhood size}
\label{ssec:av_path}



In this section we apply the Joyal's theory of combinatorial species to investigate the number of nodes that are contained within a given topological distance from a randomly chosen node.
The expected number of first-degree neighbours is defined as the sum of the neighbours reached by the in-edges and the neighbours reached by the out-edges. By using \glspl{GF} this number can be extracted from the degree distribution: 
\begin{equation}\label{eq:z1}
	\begin{split}
	z_1&=\left(\frac{\partial}{\partial x}+\frac{\partial}{\partial y}\right) U(x,y)\Big|_{x,y=1}=2\mu,
	\end{split}
\end{equation}
and moreover, the expected number of the $m^\text{th}$-degree neighbours is given by a composition of $U(x,y)$ and $m-1$-fold composition of the excess generating functions:
\begin{equation}\label{eq:zm_dir}
	z_m=\left(\frac{\partial}{\partial x}+\frac{\partial}{\partial y}\right) U^{[m]} (x,y)\Big|_{x,y=1}.
\end{equation}
Here $U^{[m]}$ generates the probability for the number of $m^\text{th}$ neighbours:
\begin{equation}\label{eq:Um_dir}
	U^{[m]}:=
	\begin{cases}
    		U(x,y),      & \quad \text{for } m=1,\\
    		U^{[m-1]}(U_\text{out}(x,y),U_\text{in}(x,y)),	& \quad \text{for } m>1. 
 	\end{cases}
\end{equation}
Therefore, for the first-degree neighbours we have $z_1=2\mu$.
For second-degree neighbours we have:
\begin{equation}\label{eq:z2}
	\begin{split}
	z_2&=\left(\frac{\partial}{\partial x}+\frac{\partial}{\partial y}\right) U(U_\text{out}(x,y),U_\text{in}(x,y))\Big|_{x,y=1}\\
	&=\mu \left(\frac{1}{\mu} \frac{\partial^2}{\partial^2 x}+2\frac{1}{\mu} \frac{\partial^2}{\partial x \partial y}+ \frac{1}{\mu} \frac{\partial^2}{\partial^2 y}\right)U(x,y)\Bigg|_{x,y=1}\\
	&= \mu \boldsymbol{1}^T \left(\frac{1}{\mu} \boldsymbol{L} U(x,y)|_{x,y=1}\right)\boldsymbol{1},
	\end{split}
\end{equation}
where $\boldsymbol{L}=\begin{pmatrix} \frac{\partial^2}{\partial x\partial y} & \frac{\partial^2}{\partial^2 y} \\ \frac{\partial^2}{\partial^2 x} & \frac{\partial^2}{\partial x\partial y}\end{pmatrix}$ and $\boldsymbol{1}^T=\begin{pmatrix}1&1\end{pmatrix}$.
The expected number of the third-degree neighbours is given by:
\begin{equation}\label{eq:z2}
	\begin{split}
	z_3&=\left(\frac{\partial}{\partial x}+\frac{\partial}{\partial y}\right) U(U_\text{out}(U_\text{out}(x,y),U_\text{in}(x,y)),U_\text{in}(U_\text{out}(x,y),U_\text{in}(x,y)))\Big|_{x,y=1}\\
	&=\mu    \Bigg( \frac{1}{\mu^2} \frac{\partial^2}{\partial x \partial y}   \left( \frac{\partial^2}{\partial x \partial y} +\frac{\partial^2}{\partial^2 y} \right)  +  \frac{1}{\mu^2} \frac{\partial^2}{\partial^2 y} \left(\frac{\partial^2}{\partial^2 x} +\frac{\partial^2}{\partial x \partial y} \right) \\
	&+ \frac{1}{\mu^2} \frac{\partial^2}{\partial^2 x}  \left(  \frac{\partial^2}{\partial x \partial y}+\frac{\partial^2}{\partial^2 y} \right)  + \frac{1}{\mu^2} \frac{\partial^2}{\partial x \partial y} \left(  \frac{\partial^2}{\partial^2 x}  +\frac{\partial^2}{\partial x \partial y}\right)\Bigg)U(x,y) \Bigg|_{x,y=1}\\
	&=  \mu \boldsymbol{1}^T \left(\frac{1}{\mu} \boldsymbol{L} U(x,y)|_{x,y=1}\right)^2 \boldsymbol{1} .
	\end{split}
\end{equation}
By using induction we arrive at the following expression for the expected number of the $m^\text{th}$-degree neighbours in terms of the degree distribution moments:
\begin{equation}\label{eq:l_dir}
	z_m=\mu \boldsymbol{1}^T \boldsymbol{A}^{m-1} \boldsymbol{1},
\end{equation}
where
\begin{equation}\label{eq:A_mu}
\boldsymbol{A}=\frac{1}{\mu}\boldsymbol{L}U(x,y)|_{x,y=1}=\frac{1}{\mu}\begin{pmatrix}  \mu_{11} &  \mu_{02}-\mu  \\ \mu_{20}-\mu  &  \mu_{11} \end{pmatrix}.
\end{equation}
Now, the total number of nodes contained within a topological ball of radius $l$ is given by as a sum:
\begin{equation}\label{eq:l_dir}
N=1+\sum_{m=1}^l z_m=	1+\mu\sum_{m=1}^{l}\boldsymbol{1}^T\boldsymbol{A}^{m-1} \boldsymbol{1}=1+\mu\boldsymbol{1}^T\left(\sum\limits_{m=1}^{l}\boldsymbol{A}^{m-1}\right) \boldsymbol{1}.
\end{equation}
Using the equation for the sum of the geometric series, $\sum_{m=0}^{l}\boldsymbol{A}^{m}=(\boldsymbol{A}^{l+1}-\boldsymbol{\boldsymbol{I}})(\boldsymbol{A}-\boldsymbol{\boldsymbol{I}})^{-1}$, with $\boldsymbol{I}$ denoting the identity matrix, we obtain:
\begin{equation}\label{eq:l_dir_nosum}
N=	1+\mu \boldsymbol{1}^T  (\boldsymbol{A}^{l}-\boldsymbol{\boldsymbol{I}})(\boldsymbol{A}-\boldsymbol{\boldsymbol{I}})^{-1} \boldsymbol{1}.
\end{equation}
The latter equality transforms to:
\begin{equation}
	\boldsymbol{1}^T  \boldsymbol{A}^{l}(\boldsymbol{A}-\boldsymbol{\boldsymbol{I}})^{-1} \boldsymbol{1} =\frac{N-1}{\mu}+\boldsymbol{1}^T (\boldsymbol{A}-\boldsymbol{\boldsymbol{I}})^{-1} \boldsymbol{1}.
\end{equation}
We will now perform an asymptotic analysis of the latter equation by assuming  that $l\gg1$, in which case,  the asymptotic behaviour of $\boldsymbol{A}^l$ is driven by the leading eigenvalue of $\boldsymbol{A}$.
Using the eigendecomposition 
$\boldsymbol{A}^l=\boldsymbol{PD}^l\boldsymbol{P}^{-1}$ gives
$	\boldsymbol{1}^T  \boldsymbol{PD}^l\boldsymbol{P}^{-1}(\boldsymbol{A}-\boldsymbol{\boldsymbol{I}})^{-1} \boldsymbol{1} =\frac{N-1}{\mu}+\boldsymbol{1}^T (\boldsymbol{A}-\boldsymbol{\boldsymbol{I}})^{-1} \boldsymbol{1},
$
and defining $\boldsymbol{a}^T=\boldsymbol{1}^T\boldsymbol{P}$, $\boldsymbol{b}=\boldsymbol{P}^{-1}(\boldsymbol{A}-\boldsymbol{I})^{-1}\boldsymbol{1}$ and $\mathcal{C}=\boldsymbol{1}^T (\boldsymbol{A}-\boldsymbol{\boldsymbol{I}})^{-1} \boldsymbol{1}$ leads to
$$	\boldsymbol{a}^T  \boldsymbol{D}^l\boldsymbol{b}=\frac{N-1}{\mu}+\mathcal{C},$$
or equivalently,
\begin{equation}\label{eq:av}
	a_1b_1\lambda_1^l+a_2b_2\lambda_2^l=\frac{N-1}{\mu}+\mathcal{C},
	\end{equation}
where the eigenvalues of the matrix $\boldsymbol{A}$, $\lambda_{1,2}=\frac{1}{\mu}\left( \mu_{11}\pm\sqrt{(\mu_{20}-\mu)(\mu_{02}-\mu)}\right)$, are defined by the characteristic polynomial: 
\begin{equation}
\left|\frac{1}{\mu}\begin{pmatrix}  \mu_{11} &  \mu_{02}-\mu  \\ \mu_{20}-\mu  &  \mu_{11} \end{pmatrix}-\lambda\boldsymbol{I}\right|=0. 
\end{equation}
Note that the gel criterion given by Eq.~(\ref{eq:gel1}) can also be rewritten as a determinant:
\begin{equation} \label{eq:gel1a}
\det \boldsymbol A'=\left|\frac{1}{\mu}\begin{pmatrix}\mu_{11}&\mu_{02}-\mu\\\mu_{20}-\mu&\mu_{11}\end{pmatrix}-\boldsymbol{I}\right|\le0.
\end{equation}
Thus, at the gel point, the matrix $\boldsymbol{A'}$ has at least one eigenvalue equal to zero. The relation of the eigenvalues of matrix $\boldsymbol{A}'$, $\lambda_1'$ and $\lambda_2'$, to the eigenvalues of $\boldsymbol{A}$, $\lambda_1$ and $\lambda_2$, is as follows: $\lambda_1'=\lambda_1-1$ and $\lambda_2'=\lambda_2-1$. Furthermore, above the gel transition, $\det \boldsymbol{A}'=\lambda_1'\lambda_2'<0.$ This is the case, only if one eigenvalue, $\lambda_1'>0$, is positive and the other one, $\lambda_2'<0$, is negative, and therefore, the eigenvalues of $\boldsymbol{A}$ satisfy $\lambda_1>1$, $\lambda_2<1$ and $|\lambda_1|>|\lambda_2|$. The implications for Eq.~(\ref{eq:av}) are as follows: for large $l$, $\lambda_1^l\gg\lambda_2^l$, and consequently, $N$ features the exponential growth after the gel transition: 
\begin{equation}\label{av}
	N \propto \lambda_1^l = e^{l/l_0},\; l_0 =(\log \lambda_1)^{-1}, \;  l\gg1.
\end{equation}
where $\lambda_1>1$ is the largest eigenvalue of matrix $\boldsymbol A$. 

\section{Acknowledgment}
This research was supported by Oc\'{e} Technologies B.V. and the Technology Foundation STW. IK acknowledges support form research program Veni with project number 639.071.511, which is financed by the \gls{NWO}.

\section*{Competing interests}
The authors declare no competing interests.

\section*{Author Contributions}
IK designed research and developed the theoretical formalism, VS derived analytical equations, performed numerical simulations, and composed figures, VS and PDI performed the literature study, wrote Introduction and Results, VS and IK wrote Methods. All authors discussed the results and contributed to writing the final manuscript.
\bibliography{literature}
\end{document}